\documentclass[11pt]{article}
\usepackage{graphicx}
\usepackage[margin=1in]{geometry}
\usepackage{authblk}
\usepackage[numbers,sort&compress]{natbib}

\title{Low-dimensional structure and online tracking of POD subspaces on the Grassmann manifold: application to flow around an airfoil}
\author[1]{Shintaro Sato}
\author[2]{Yoshitsugu Naka}
\author[1]{Rei Sasaki}
\author[2]{Rion Handa}
\author[1]{Naofumi Ohnishi}
\date{}

\affil[1]{Department of Aerospace Engineering, Tohoku University, Aramaki-aza-Aoba 6-6-01, Aoba-ku, Sendai 980-8579, Japan}
\affil[2]{Department of Mechanical Engineering, Meiji University, Kawasaki, 214-8571, Japan}

\begin{document}

\maketitle

\begin{abstract}
  Representing flow states over a wide range of flow parameters and control inputs in a low-dimensional state space is a central challenge in fluid mechanics.
  Rather than representing the instantaneous flow field in a fixed subspace spanned by the leading proper orthogonal decomposition (POD) modes,
  this study regards the POD subspace itself as the flow state at each flow condition.
  The set of POD subspaces associated with flow conditions defines a state space on the Grassmann manifold.
  Diffusion maps identify the intrinsic low-dimensional structure of the family of POD subspaces,
  while Grassmannian rank-one update subspace estimation tracks the temporal evolution of a POD subspace online.
  The framework is experimentally demonstrated for flow around an airfoil.
  POD subspaces are extracted from wall-pressure fluctuations measured by a microphone array over a range of angles of attack and under different control inputs.
  The subspaces are found to lie on a one-dimensional submanifold of the Grassmann manifold.
  Moreover, the transition from separated to attached flow, induced by a plasma actuator, follows a reproducible trajectory along the same submanifold identified from statistically stationary flow data.
  The temporal evolution of the subspace is consistent with the transient evolution of the flow field observed using particle image velocimetry.
  These results show that POD subspaces can serve as representative flow states,
  enabling their temporal evolution across a wide range of flow conditions to be tracked online in a low-dimensional space based on wall-pressure fluctuations.
  This low-dimensional representation provides a basis for real-time flow-state estimation and feedback control.  
\end{abstract}

\section{Introduction}
A fundamental challenge in fluid mechanics, particularly in engineering applications, is to identify and represent the state of complex turbulent flows
from limited observations so that the future state can be predicted and the flow can be guided toward a more desirable state.
Turbulent flows are inherently high-dimensional and multiscale,
making fully resolved state estimation from limited observations and real-time prediction extremely challenging. 
For active flow control, however, resolving a high-dimensional full flow state is not always necessary.
Instead, previous studies have sought to construct reduced representations of the flow state that capture the dominant flow structures and to develop reduced-order models (ROMs)
for predicting the temporal evolution of the reduced state and determining an appropriate control input\cite{Gillies1998, Semeraro2011, Deem2020}.

Modal analysis provides a well-established framework for extracting dominant flow structures from time-resolved flow-field data obtained
from experiments or numerical simulations\cite{Taira2017}.
ROMs based on modal analysis have been successfully applied to active flow control\cite{Noack2011}.
However, because the dominant modes can vary with flow parameters and control inputs,
a fixed reduced subspace constructed for a specific condition may fail to represent the dominant flow structures and dynamics under different conditions\cite{Benner2015}.
Machine learning has recently been used to represent flow dynamics across multiple parameter values on a low-dimensional manifold\cite{Fukami2023}.
Nevertheless, representing flow states over a wide range of flow parameters and control inputs in a fixed reduced state space remains a major challenge.

In this study, rather than representing an instantaneous flow field by coordinates in a fixed subspace spanned by proper orthogonal decomposition (POD) modes,
we regard the POD subspace associated with each flow condition as a representation of the flow state.
This viewpoint is motivated by active-flow-control problems in which the objective is often to modify the dominant flow structures
rather than to track a prescribed trajectory of the fully resolved flow field.
Therefore, the collection of POD subspaces associated with different flow conditions can be regarded as a state space.
The set of all $r$-dimensional subspaces of an $n$-dimensional vector space forms the Grassmann manifold $\mathrm{Gr}(n,r)$\cite{Edelman1998},
providing a natural geometric space for this state representation.
Each POD subspace is represented as a point on the Grassmann manifold while
the temporal evolution of the POD subspace during a flow-state transition is represented as a trajectory on the manifold.

Although the dimension of $\mathrm{Gr}(n,r)$ is $r(n-r)$, which is generally large in fluid-mechanical applications because the dimension $n$ is typically high,
the family of POD subspaces obtained over a wide range of flow parameters and control inputs is expected to form a structure of much lower intrinsic dimension
within the Grassmann manifold
because these POD subspaces are constrained by the underlying flow physics and vary with flow parameters rather than being arbitrary.
A previous study showed that the geometric characteristics of a curve on the Grassmann manifold that represents the dependence of POD subspaces on a flow parameter 
are closely related to the fluid-mechanical characteristics of flows represented by those subspaces\cite{Sato2025a}.

This study applies the subspace-based state representation to the flow over an airfoil,
using the spatial distribution of wall-pressure fluctuations measured by a microphone array as a sensor-based observable of the flow field.
We show that the family of POD subspaces obtained over a range of angles of attack and control parameters forms a one-dimensional structure on the Grassmann manifold.
We further demonstrate that, during a transition from a separated to an attached state induced by the operation of an active flow-control device,
the temporal evolution of the subspace traces a trajectory along the one-dimensional structure and can be tracked online from the sensor measurements.
These results highlight that the family of POD subspaces provides a reduced state space for representing flow states over a wide range of flow conditions
and a foundation for models that forecast future states and determine appropriate control inputs.

\section{Methods}

\begin{figure}
  \begin{center}
    \includegraphics[width=1.0\textwidth]{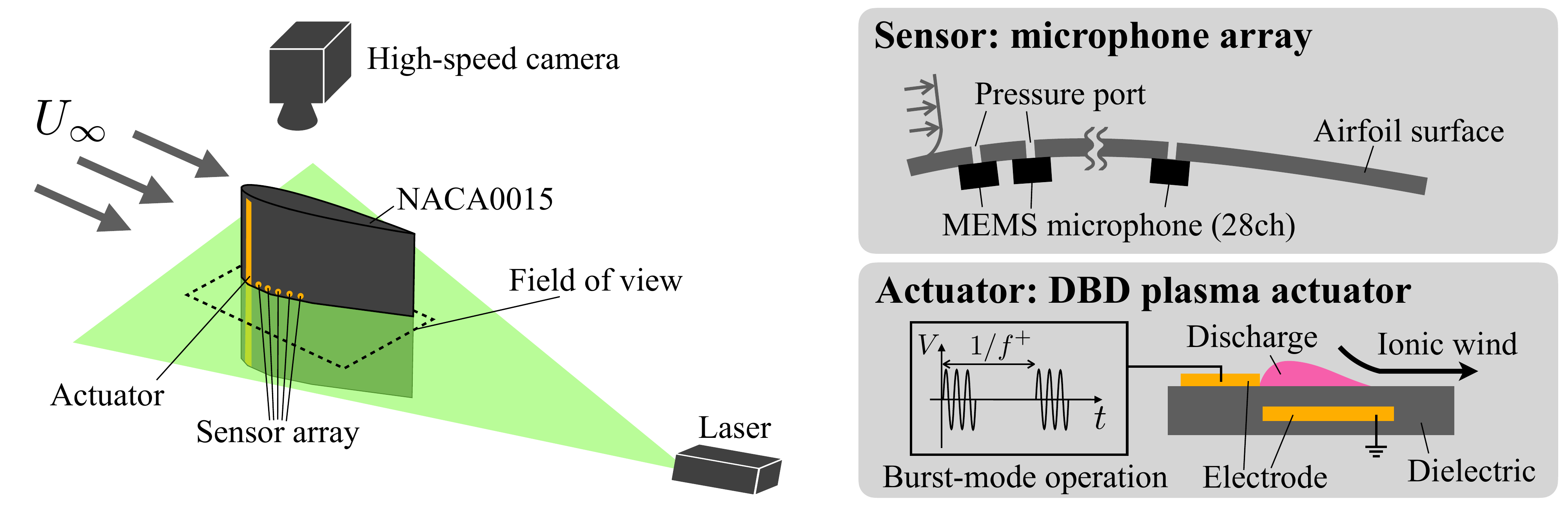}
    \caption{Schematic of experimental setup for simultaneous measurement of velocity field and wall-pressure fluctuations around an airfoil controlled by a DBD plasma actuator.\label{fig:figure1}}
  \end{center}
\end{figure}
The schematic experimental setup is shown in figure~\ref{fig:figure1}.
The experiments were carried out in a blowing-type wind tunnel with a nozzle exit of $400~\mathrm{mm}\times400~\mathrm{mm}$.
The free-stream velocity was fixed at $4~\mathrm{m/s}$ for all experiments.
A NACA0015 airfoil model with a chord length of $400~\mathrm{mm}$ and span length of $400~\mathrm{mm}$ was employed,
resulting in a chord-based Reynolds number of approximately $1.06\times10^5$.

A dielectric-barrier-discharge (DBD) plasma actuator is installed on the suction side at $x/c=0.05$,
where $x$ denotes the chordwise distance from the leading edge and $c$ is the chord length.
The DBD plasma actuator consists of two electrodes and a dielectric layer and adds momentum to the flow field through electrohydrodynamic force that acts as a body force\cite{Sato2025b}.
Copper tape with a thickness of $0.1~\mathrm{mm}$ and a width of $10~\mathrm{mm}$ was used for both the exposed and embedded electrodes.
The dielectric layer consisted of polyimide tape with a thickness of $0.09~\mathrm{mm}$ and a width of $10~\mathrm{mm}$.
The spanwise length of the DBD plasma actuator was $300~\mathrm{mm}$.
The surface DBD was generated by applying an alternating-current (AC) voltage waveform with an amplitude of $6~\mathrm{kV}$ and a frequency of $10~\mathrm{kHz}$
to the exposed electrode using a high-voltage power supply (PSI Inc. PG1040F).
Burst-mode operation, in which an AC voltage is applied intermittently rather than continuously,
was employed to suppress the flow separation\cite{Huang2006} (figure~\ref{fig:figure1}).
The nondimensional burst frequency is defined as $F^+=f^+c/U_\infty$, where $f^+$ denotes the burst frequency and $U_\infty$ is the freestream velocity.
We performed experiments at $F^+=0.1$, $1$, and $10$.

Active flow control framework requires sensors to obtain information on the flow state in addition to an actuator.
Motivated by the practical implementation, we employ an array of miniature microphone mounted on the airfoil surface
to obtain the spatial distribution of the wall-pressure fluctuations.
A previous study showed that the spatial and temporal variations of wall-pressure fluctuations on an airfoil surface are closely related to
the development and convection of large-scale vortical structures in the separated shear layer\cite{Sen2023}.
Therefore, wall-pressure measurements are expected to provide useful information for identifying changes in the flow state around the airfoil.
A total of 28 digital MEMS microphones (INMP621, TDK InvenSnse) were installed over the chordwise range $0.125\le x/c\le0.395$.
The center-to-center spacing of adjacent pressure ports was $4~\mathrm{mm}$.
The port diameter was $0.5~\mathrm{mm}$.
The microphone array was flush-mounted on the wing model surface at mid-span.
The pressure fluctuations were sampled with a frequency of $25~\mathrm{kHz}$.
The sampled waveforms were converted to pascals using a calibration factor obtained by acoustic calibration against a reference microphone
(Br\"uel \& Kj\ae r, 1/4-inch, Type 4938-A-011).

In addition to the wall-pressure-fluctuation measurements, particle image velocimetry (PIV) was performed simultaneously to obtain time-resolved flow-field data.
The velocity fields obtained by PIV were used solely for visualization and were not used for flow-state estimation.
The laser beam from a high-repetition-rate dual-cavity laser system (DM10, Photonics Industries) was shaped into a thin light sheet.
The light sheet was oriented in a chordwise-normal plane and positioned close to the pressure ports of the microphone array.
The field of view was approximately $380~\mathrm{mm}\times 210~\mathrm{mm}$.
The high-speed camera (Phantom VEO610L, Vision Research) was operated at a frame rate of $2~\mathrm{kHz}$.

\section{Relation between the flow field and surface-pressure POD modes}\label{Sec:Relation}

\begin{figure}
  \begin{center}
    \includegraphics[width=1.0\textwidth]{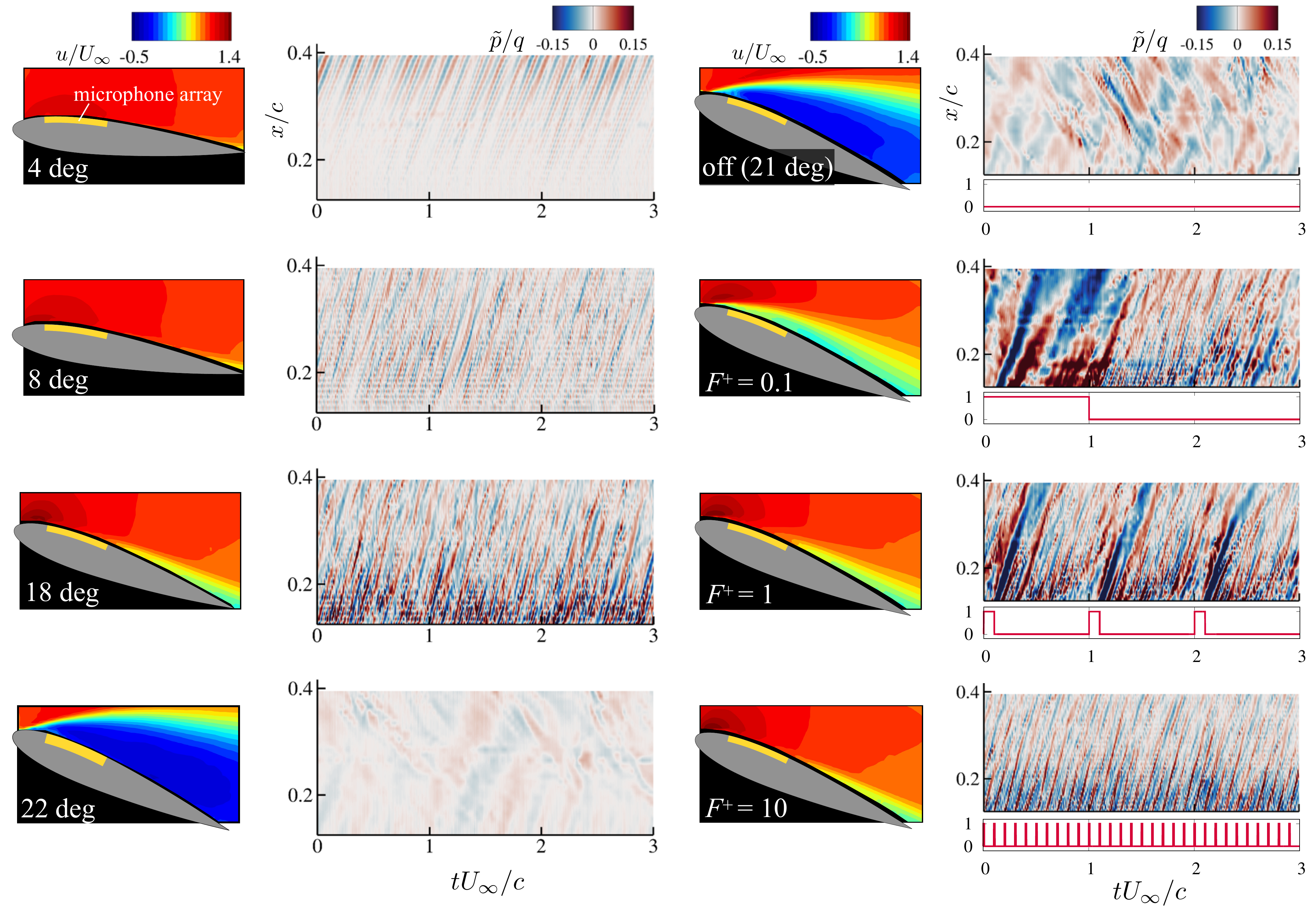}
    \caption{Time-averaged velocity fields and spatiotemporal distributions of wall-pressure fluctuations at angles of attack of 4, 8, 18, and 22 deg.
        At $21~\mathrm{deg}$, the results are shown for different operating conditions of the DBD plasma actuator (off, $F^+=0.1,1,10$).
        The actuation signal is also shown, with a value of 1 indicating that the actuator is on.
        The wall-pressure fluctuations are normalized by the dynamic pressure $q$.
        \label{fig:figure2}}
  \end{center}
\end{figure}
Wind-tunnel experiments were first carried out without operating the DBD plasma actuator at angles of attack ranging from $4~\mathrm{deg}$ to $22~\mathrm{deg}$
in increments of $2~\mathrm{deg}$.
The spatial distributions of the time-averaged streamwise velocity and the spatiotemporal distributions of wall-pressure fluctuation are shown in figure~\ref{fig:figure2}.
The time-averaged velocity fields indicate that the flow is attached to the airfoil at angles of attack up to $18~\mathrm{deg}$,
whereas the flow is separated at $\alpha\ge 20~\mathrm{deg}$.
The spatiotemporal distributions of wall-pressure fluctuations show that,
at $4~\mathrm{deg}$, relatively weak fluctuations occur mainly in the downstream region $(x/c>0.3)$.
Within the attached-flow regime, the region of large pressure fluctuations extends upstream and their amplitude increases with the angle of attack.
This upstream shift is consistent with a previous study\cite{Boutilier2012}, which showed that the maximum pressure-fluctuation amplitude occurs between
the laminar-to-turbulent transition and reattachment points of the laminar separation bubble
and moves toward the leading edge with increasing angle of attack.
Moreover, when the flow is attached, the pressure fluctuation structures propagate downstream at a speed comparable to the freestream velocity.
In contrast, once the flow becomes fully separated, the amplitude of the wall-pressure fluctuations decreases substantially.
Additionally, not only the pressure fluctuations propagating downstream, but also those propagating upstream observed,
suggesting the presence of a recirculation region above the airfoil surface.

At $21~\mathrm{deg}$, the flow is fully separated without operation of the DBD plasma actuator,
and the pressure fluctuation pattern is similar to that at $22~\mathrm{deg}$.
At $F^+=0.1$, the time-averaged velocity field indicates that the recirculation region is reduced, although the instantaneous flow exhibits repetitive transition
between separated and attached states, as discussed later.
For $1<tU_\infty/c<2$, the pressure fluctuation pattern is similar to that at $18~\mathrm{deg}$,
suggesting that the flow is attached.
The flow then gradually separates, resulting in a spatiotemporal pattern of wall-pressure fluctuations similar to that observed without control.
At $F^+=1$ and $10$, the flow remains attached throughout the measurement.
As $F^+$ increases, the spatiotemporal distribution of wall-pressure fluctuations becomes more similar to that observed at an angle of attack of $18~\mathrm{deg}$.
The dependence of the pressure-fluctuation patterns on the angle of attack and control input indicates that
the dominant flow characteristics can be identified from these patterns.

\begin{figure}
  \begin{center}
    \includegraphics[width=1.0\textwidth]{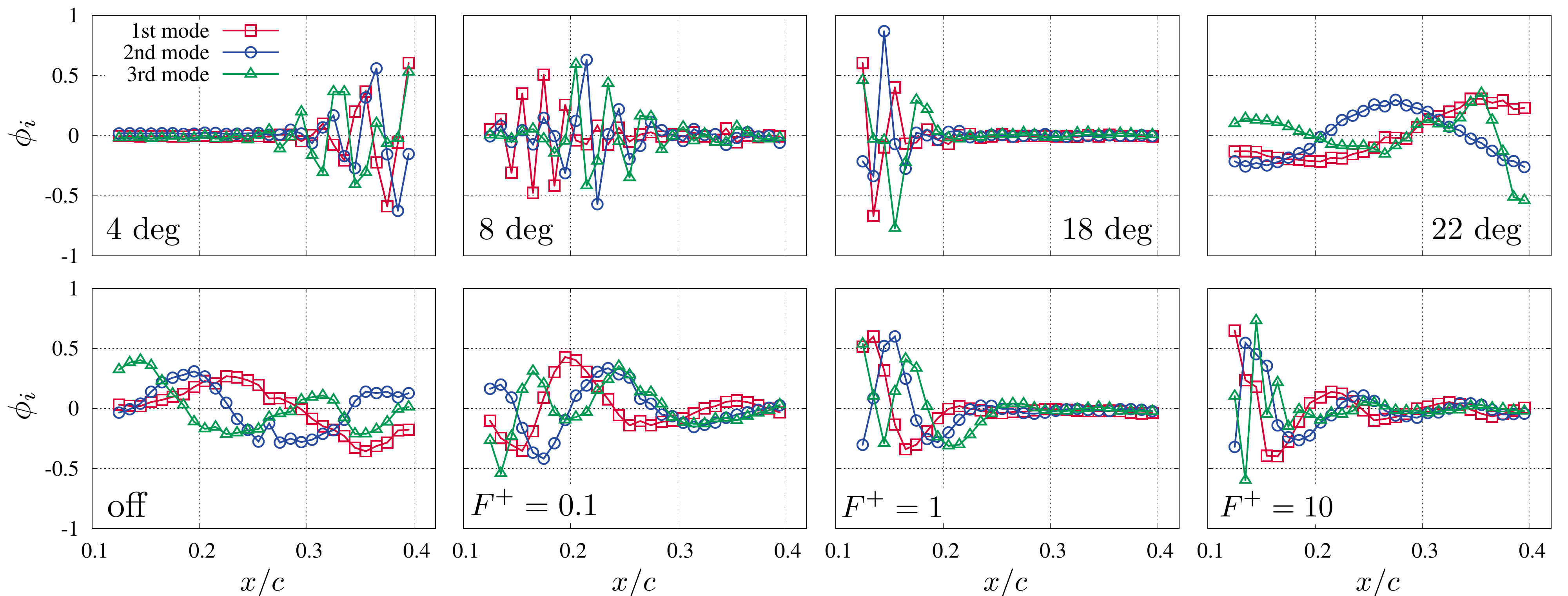}
    \caption{Spatial distributions of leading three POD modes extracted from wall-pressure fluctuations at angles of attack 4, 8, 18, and 22 deg.
      At $21~\mathrm{deg}$, the results are shown for different operating conditions of the DBD plasma actuator (off, $F^+=0.1,1,10$).\label{fig:figure3}}
  \end{center}
\end{figure}
POD analysis is then performed on a snapshot matrix constructed from time-series data of wall-pressure fluctuations to extract the dominant structures.
The snapshot matrix consists of 25,000 snapshots sampled over $tU_\infty/c=10$ at a sampling interval of $tU_\infty/c=4\times 10^{-4}$.
Figure~\ref{fig:figure3} shows the leading three POD modes for the flow conditions shown in figure~\ref{fig:figure2}.
For attached flows, the region of large modal amplitude shifts upstream with increasing angle of attack.
The separated-flow cases at $22~\mathrm{deg}$ and at $21~\mathrm{deg}$ without control exhibit larger-scale spatial structures,
whereas the controlled cases ($F^+=0.1,1$, and 10) at $21~\mathrm{deg}$ exhibit localized structures near the leading edge
similar to those at $18~\mathrm{deg}$.
These results show the same trends as those observed in the spatiotemporal distributions of wall-pressure fluctuations
and indicate that variations in flow characteristics with angle of attack and control input are reflected in the dominant POD modes,
motivating the use of the POD subspace as a representation of the flow state.

\section{Low-dimensional structure of family of POD subspaces on the Grassmann manifold}\label{Sec:Low-dim}
To identify the intrinsic low-dimensional structure formed by POD subspaces obtained for various angles of attack and the control parameter of the DBD plasma actuator,
we apply diffusion maps\cite{Coifman2005}, which is a manifold learning method.
It should be noted that the manifold considered in manifold learning is different from the Grassmann manifold itself.
In this study, manifold learning is used to reveal a lower-dimensional submanifold formed by the family of POD subspaces within the Grassmann manifold.

This study regards each POD subspace as a point on the Grassmann manifold.
In diffusion maps, the subspaces sampled from experimental measurements are used as the nodes of a weighted graph.
The edge weight between two subspaces, $\mathcal{S}_i$ and $\mathcal{S}_j$, is defined from their distance on the Grassmann manifold using the following kernel:
\begin{equation}
  k(\mathcal{S}_i,\mathcal{S}_j) = \exp\left[ -\frac{d^2_\mathrm{Gr}(\mathcal{S}_i,\mathcal{S}_j)}{\varepsilon} \right],
\end{equation}
where $d_\mathrm{Gr}$ denotes the distance between two subspaces on the Grassmann manifold and the scale parameter $\varepsilon$ is set to 2.4.
In this study, the geodesic distance is used.
For details on the representation of a POD-mode matrix as a point on the Grassmann manifold and the computation of the distance,
we refer to Ref.\cite{Sato2025a}.
The kernel matrix $K$, whose elements are $K_{ij}=k(\mathcal{S}_i,\mathcal{S}_j)$, is normalized as
$\widetilde{K}_{ij}=\frac{K_{ij}}{q_iq_j}$, where $q_i=\sum_j K_{ij}$.
The Markov transition matrix is then obtained as follows:
\begin{equation}
  P_{ij}=\frac{\widetilde{K}_{ij}}
  {\sum_j\widetilde{K}_{ij}}.
\end{equation}
The leading non-trivial right eigenvectors of $P$ provide a low-dimensional representation of subspace distribution, referred to as diffusion coordinates\cite{Coifman2005}.

\begin{figure}
  \begin{center}
    \includegraphics[width=0.6\textwidth]{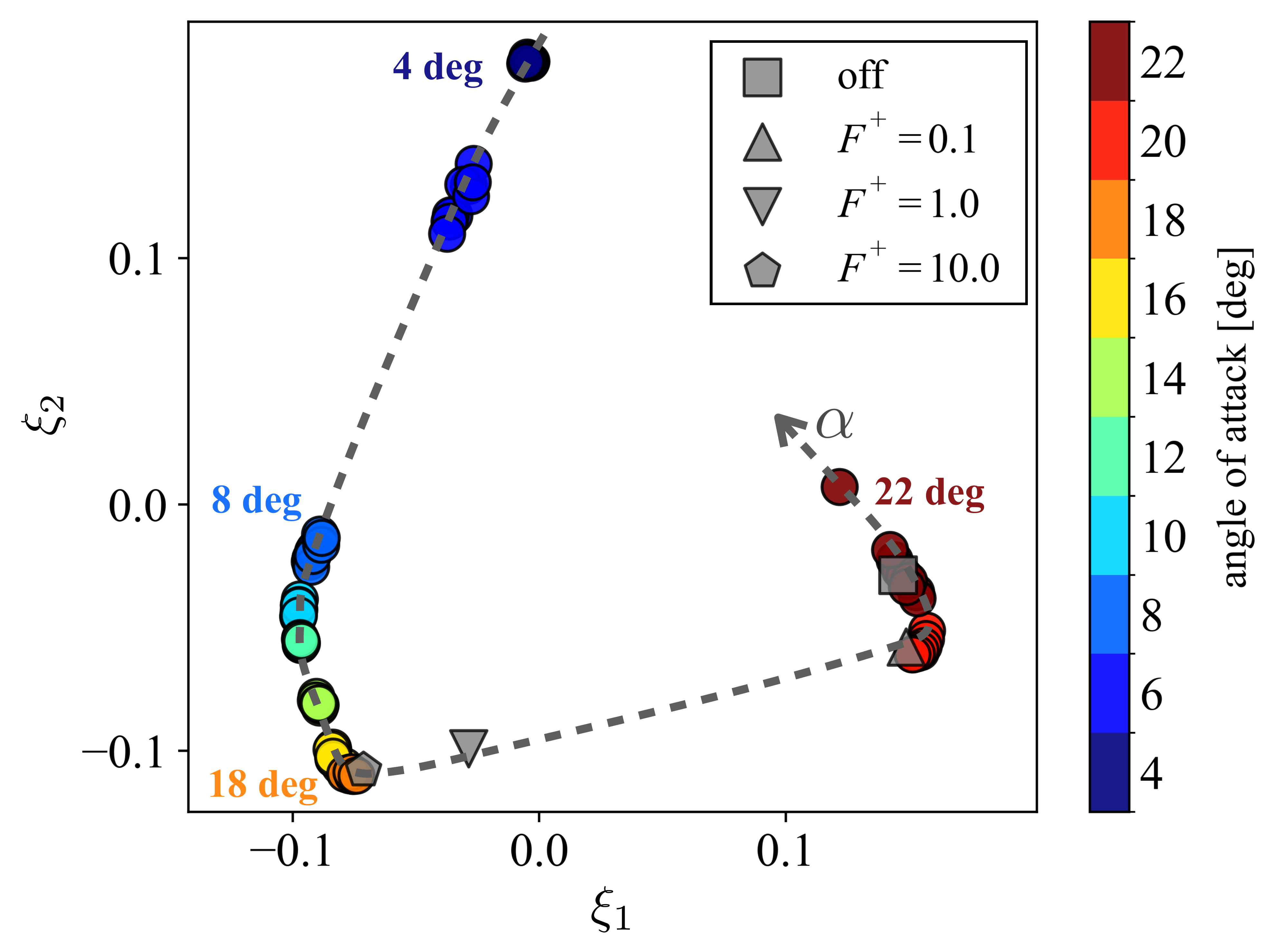}
    \caption{Diffusion map embedding of the POD subspaces using the first two non-trivial diffusion coordinates $\xi_1$ and $\xi_2$.
      POD subspaces at $21~\mathrm{deg}$ for different operating conditions of the DBD plasma actuator are mapped via the Nystr\"om extension.
      \label{fig:figure4}}
  \end{center}
\end{figure}
Figure~\ref{fig:figure4} shows the distribution of the POD subspaces obtained in this study in the first two non-trivial diffusion coordinates.
The training data consist of nine POD subspaces at each angle of attack from $4~\mathrm{deg}$ to $22~\mathrm{deg}$ in $2~\mathrm{deg}$ increments.
At each angle of attack, nine POD subspaces were obtained from nine separate measurement snapshot data.
Note that the POD subspaces at $21~\mathrm{deg}$, both without and with DBD plasma actuator operation, were excluded from the training data
to evaluate the validity of the identified low-dimensional structure.
The POD subspaces lie along an approximately one-dimensional structure and vary continuously with angle of attack,
although angle-of-attack information was not used to construct the diffusion map.
This result indicates that the variation of the dominant flow structures with angle of attack forms an intrinsically one-dimensional geometric structure
on the Grassmann manifold.
Furthermore, the POD subspaces for attached and separated flows lie in clearly distinct regions of the embedding space obtained by the diffusion map,
reflecting the large difference in the dominant flow structures between attached and separated flows.

To further examine the physical relevance of the identified one-dimensional structure,
the POD subspaces that were not used to construct the diffusion map, i.e., those at $21~\mathrm{deg}$, were mapped into the embedding space
using the Nystr\"om extension, without recomputing the eigenvalues and eigenvectors\cite{Coifman2008}.
These out-of-sample POD subspaces also lie on the one-dimensional structure identified from the training data.
The POD subspace at $21~\mathrm{deg}$ without operation of the DBD plasma actuator is located in the region where the subspaces corresponding to separated flows are distributed.
In contrast, with increasing $F^+$, the POD subspaces shift along the one-dimensional structure toward the region associated with attached flows at lower angles of attack.
This trend is consistent with the PIV measurements.
Therefore, changes in the dominant flow structures caused by the DBD plasma actuator are represented as
shifts of the POD subspaces along the one-dimensional structure parametrized by angle of attack.

\section{Online tracking of time evolution of POD subspaces during a flow-state transition}
In \S\ref{Sec:Low-dim}, we analyzed the low-dimensional structure of the family of POD subspaces extracted from snapshot data of statistically stationary flows.
To analyze the temporal evolution of the flow state during a transient process using our framework, subspace tracking\cite{Balzano2018} is introduced.
Subspace tracking recursively updates a subspace that characterizes the dominant flow structures using time-series measurements.

In this study, the Grassmannian Rank-One Update Subspace Estimation (GROUSE) algorithm
\cite{Balzano2010} is employed to track the subspace.
GROUSE updates the current subspace by performing gradient descent on the Grassmann manifold to reduce the reconstruction error of each newly obtained snapshot.
The update is based on the projection residual.
In other words, when the snapshot is not well represented by the current space, the space is modified in the direction indicated by the residual.
Conversely, when the reconstruction error is small, little update is made. 
Details of the computational procedure of GROUSE are given in Refs.\cite{Balzano2010,Balzano2018}.

Subspace tracking provides a dynamical-system interpretation of the flow state on the Grassmann manifold.
For a statistically stationary flow, the subspace representing the dominant flow structures is expected to remain unchanged;
therefore, sequentially acquired snapshots only produce small changes in the updated subspace.
Ideally, a statistically stationary flow is thus represented as a fixed point on the Grassmann manifold.
In contrast, when a disturbance or control input modifies the dominant flow structures,
the temporal evolution of the subspace during the transient process is represented as a trajectory on the Grassmann manifold.

\begin{figure}
  \begin{center}
    \includegraphics[width=0.8\textwidth]{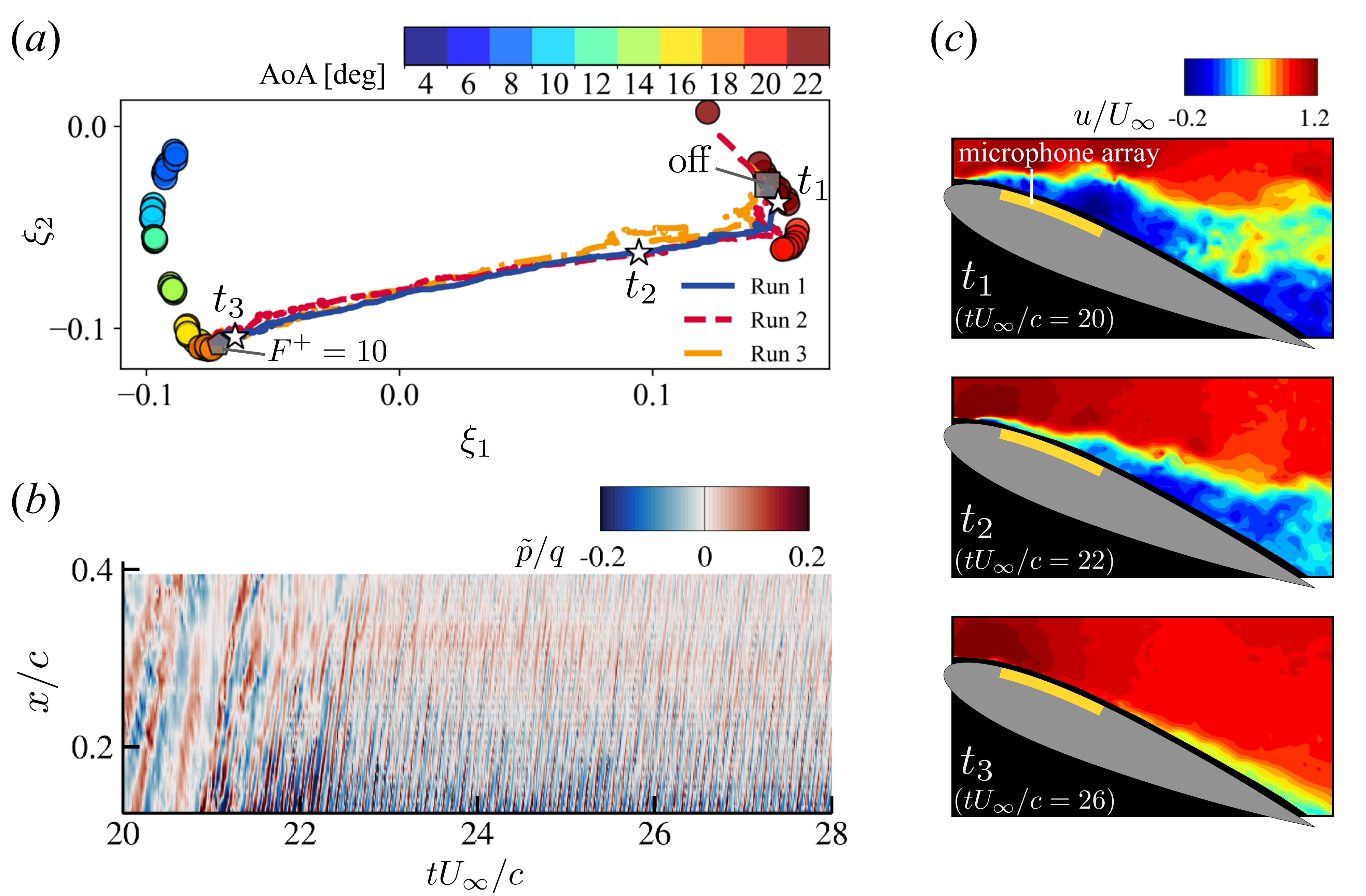}
    \caption{Temporal evolution of flow state induced by operation of the DBD plasma actuator at $F^+=10$. (a) POD subspace trajectories.
      (b) Spatiotemporal distribution of wall-pressure fluctuations for Run 1.
      (c) Velocity fields at $t_1$, $t_2$, and $t_3$, marked on the Run 1 trajectory in (a).\label{fig:figure5}}
  \end{center}
\end{figure}
Figure~\ref{fig:figure5} (a) shows the trajectories of the subspace during the transition from separated to attached flow
induced by the operation of the DBD plasma actuator at $F^+=10$ for an angle of attack of $21~\mathrm{deg}$.
The three trajectories were obtained from three experiments under the same experimental conditions.
The subspaces tracked by GROUSE were mapped into the two-dimensional diffusion-coordinate space using the Nystr\"om extension.
The three trajectories follow a reproducible path along the one-dimensional structure identified in \S\ref{Sec:Low-dim}.
Each trajectory starts from the POD subspace representing the flow at $21~\mathrm{deg}$ without control
and approaches that representing the flow with the DBD plasma actuator operated at $F^+=10$.
These results show that, despite the complexity of the instantaneous flow field,
the transient process can be represented by a simple subspace trajectory on the Grassmann manifold.
Furthermore, the fact that the trajectories during the transient process follow the one-dimensional structure identified from data for statistically stationary flows suggests
that the temporal evolution of the subspace in response to the control input is constrained to the same structure parametrized by angle of attack.
The spatiotemporal distribution of the wall-pressure fluctuations during the transient process (figure~\ref{fig:figure5}b) varies from a relatively large-scale pattern characteristic of separated flow
to a finer-scale pattern associated with attached flow.
The velocity fields measured by PIV (figure~\ref{fig:figure5}c) confirm that this transition in the wall-pressure pattern
corresponds to the transition from separated to attached flow.
Consequently, the subspace trajectory captures the change in the wall-pressure fluctuation pattern and provides a low-dimensional representation of the flow field transition.

\begin{figure}
  \begin{center}
    \includegraphics[width=0.8\textwidth]{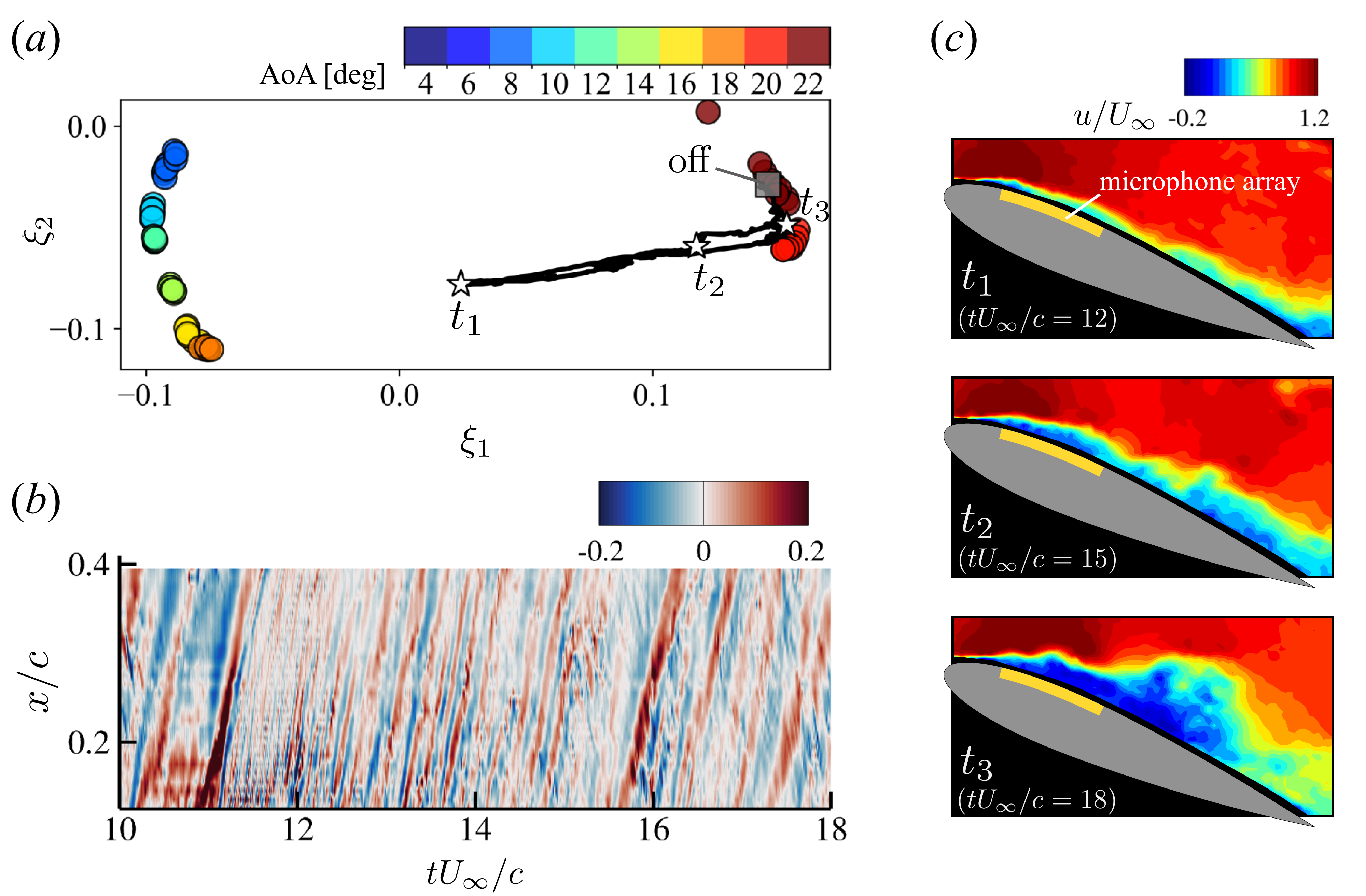}
    \caption{Temporal evolution of flow state induced by operation of the DBD plasma actuator at $F^+=0.1$. (a) POD subspace trajectory.
      (b) Spatiotemporal distribution of wall-pressure fluctuations.
      (c) Velocity fields at $t_1$, $t_2$, and $t_3$, marked on the trajectory in (a).\label{fig:figure6}}
  \end{center}
\end{figure}
The trajectory of the subspace for $F^+=0.1$ is shown in figure~\ref{fig:figure6} (a).
At $F^+=0.1$, the flow field exhibits repeated transitions between attached and separated states.
The subspace trajectory first shifts from the separated-flow region toward the attached-flow region
while the DBD plasma actuator is operated, and then returns toward the separated-flow region after the actuator is turned off.
The spatiotemporal distribution of the wall-pressure fluctuations (figure~\ref{fig:figure6}b) and
the velocity fields (figure~\ref{fig:figure6}c) also show that the flow temporarily attaches to the airfoil during operation of the DBD plasma actuator
and then gradually separates again.
Moreover, the time at which the subspace trajectory is closest to the attached-flow region
coincides with temporary attachment of the flow observed in the PIV measurements.
This correspondence indicates that the subspace trajectory in the low-dimensional space reflects the actual flow state during the transient process.
Therefore, our framework, consisting of online subspace tracking and manifold learning,
enables the temporal evolution of the flow state during a transient process to be represented as a trajectory on a low-dimensional structure.

\section{Concluding remarks}
This study showed that the family of POD subspaces identified from wall-pressure fluctuations measured on the airfoil surface
provides a representation of the dominant flow characteristics over a range of angles of attack and control inputs.
These subspaces lie on a one-dimensional submanifold within the Grassmann manifold.
The POD subspace varies continuously along this one-dimensional structure as the angle of attack changes.
Furthermore, the POD subspaces identified when the DBD plasma actuator was operated at different burst frequencies lie on the same submanifold.
These results suggest that the family of POD subspaces representing the flow around the airfoil is not distributed arbitrarily on the Grassmann manifold
but is constrained to a low-dimensional submanifold.

We also showed that GROUSE enables the temporal evolution of the POD subspace to be tracked online.
During the transient process of the flow controlled by the DBD plasma actuator,
the POD subspace evolves along the one-dimensional submanifold identified from data for statistically stationary flows.
Moreover, both a transition from one statistically stationary state to another and
a response in which the flow becomes temporarily attached and then separates again can be represented as trajectories along the same submanifold.
The temporal evolution of the subspace tracked by GROUSE is consistent with the transient evolution of the flow field observed using PIV.
These results demonstrate that online subspace tracking based on wall-pressure fluctuations captures transient flow-state responses
as trajectories on a low-dimensional submanifold of the Grassmann manifold.

Importantly, the subspace trajectories corresponding to the flow-state transitions were obtained online using only wall-pressure fluctuations measured on the airfoil,
without using velocity-field data.
This result demonstrates that the developed framework can estimate the flow state from limited surface measurements and
track its temporal evolution in a low-dimensional state space that accommodates flow states over a wide range of flow conditions.

In turbulent flows, the temporal evolution of the instantaneous flow field generally forms a complex trajectory in a high-dimensional state space.
By contrast, with a POD-subspace representation, statistically stationary flow states correspond to fixed points,
whereas transient process induced by disturbances or control inputs follow trajectories on a low-dimensional submanifold of the Grassmann manifold.
In future, modeling the temporal evolution of the POD subspace as a dynamical system on the low-dimensional submanifold leads to a practical framework for active flow control,
in which the state is guided toward a POD subspace corresponding to a desirable flow state,
rather than designing a control strategy in a state space defined by the instantaneous flow field.

\section*{Acknowledgements}
This work is supported in part by Japan Society for the Promotion of Science (JSPS) KAKENHI Grant Numbers 24K17442, 24K01073, Japan Science and Technology Agency (JST) PREST Grant Numbers JPMJPR21O4 and JPMJPR23O9.

\section*{Declaration of interests}
The authors report no conflict of interest.

\end{document}